\documentclass[10pt,letterpaper]{article}
\pdfoutput=1

\usepackage[utf8x]{inputenc}
\usepackage{hyperref}
\usepackage{amssymb}
\usepackage{amsmath}
\usepackage{siunitx}
\usepackage{graphicx}
\usepackage{subfigure}
\usepackage{xcolor}

\begin{document}
	
\begin{flushleft}
	{\Large
		\textbf\newline{The MEV project: design and testing of a new high-resolution telescope for Muography of Etna Volcano}
	}
\newline
\\
D. Lo Presti\textsuperscript{1,2},
G. Gallo\textsuperscript{1,3,*},
D.L. Bonanno\textsuperscript{2},
G. Bonanno\textsuperscript{4},\\
D.G. Bongiovanni\textsuperscript{3},
D. Carbone\textsuperscript{5},
C. Ferlito\textsuperscript{6},
J. Immè\textsuperscript{1},
P. La Rocca\textsuperscript{1,2},\\
F. Longhitano\textsuperscript{2},
A. Messina\textsuperscript{5},
S. Reito\textsuperscript{2},
F. Riggi\textsuperscript{1,2},
G. Russo\textsuperscript{1,2},\\
L. Zuccarello\textsuperscript{7,5}
\\
\bigskip
1 Department of Physics and Astronomy, University of Catania, Italy
\\
2 INFN, Sezione di Catania, Catania, Italy
\\
3 INFN, Laboratori Nazionali del Sud, Catania, Italy
\\
4 INAF, Osservatorio Astrofisico di Catania, Italy
\\
5 INGV, Sezione di Catania - Osservatorio Etneo, Catania, Italy
\\
6 Department of Biological, Geological and Environmental Sciences, University of Catania, Italy
\\
7 Universidad de Granada, Dpto. de Teoría de la Señal, Telemática y Comunicaciones ETSI Informática y de Telecomunicación Universidad de Granada, Spain
\\
\bigskip
* giuseppe.gallo@ct.infn.it

\end{flushleft}

\section*{Abstract}
The MEV project aims at developing a muon telescope expressly designed for the muography of Etna Volcano. In particular, one of the active craters in the summit area of the volcano would be a suitable target for this experiment. A muon tracking telescope with high imaging resolution was built and tested during 2017. The telescope is a tracker based on extruded scintillating bars with WLS fibres and featuring an innovative read-out architecture. It is composed of three XY planes with a sensitive area of \SI{1}{m^2}; the angular resolution does not exceeds \SI{0.4}{\milli\steradian} and the total angular aperture is about $\pm$\SI{45}{\degree}.  A special effort concerned the design of mechanics and electronics in order to meet the requirements of a detector capable to work in a hostile environment such as the top of a tall volcano, at a far distance from any facility. The test phase started in January 2017 and ended successfully at the end of July 2017. An extinct volcanic crater (the Monti Rossi, in the village of Nicolosi, about 15km from Catania) is the target of the measurement. The detector acquired data for about 120 days and the preliminary results are reported in this work. 

\subsection*{Keywords}
Muography, Experimental particle physics, Tracking detector, Geophysics, Volcanology, Natural hazard.

\section{Introduction}
Volcanic activity is regulated by complex coupled phenomena occurring in the plumbing system, from shallow depths to several kilometres below the ground surface. In persistently active volcanoes (like Etna), physical-chemical processes in the plumbing system govern an overall state of dynamical equilibrium \cite{Allard2006,FERLITO201814}. Paroxysmal phases of the activity occur when the system is driven out of its dynamical equilibrium and large amounts of volcanic material is emitted either effusively or explosively \cite{Gonnermann2007}. Mitigating volcanic hazards relies on the capability to interpret surface observations in terms of subsurface processes that may tip a volcano out of its equilibrium state \cite{Gilbert2008}. To monitor active volcanoes and also to investigate their internal structure, different techniques are utilized, mainly based on the study of round deformation, seismicity, gas emission and gravity \cite{dzurisin2006volcano,ChouetMatoza2013,Carbone2017}. Nevertheless, while reliable information can be obtained on processes that occur at greater depths, the shallowest levels of volcanic plumbing system are often poorly understood, due to the difficulty of applying classical geophysical methods to highly fractured layers, especially in the presence of steep topography.
This blind spot represents a vulnerability of volcano monitoring systems. Indeed, the modifications that occur in the shallowest level of the feeding system are not irrelevant for the analysis of volcanic hazard. It is, therefore, of great importance to develop new methods that can provide reliable information on the topmost layer of a volcano feeding system.

Muography - i.e. muon radiography - is a promising technique to address this issue. It aims at resolving the internal structure of large size target by taking advantage of the high penetrating power of cosmic muons. The basic properties of muon interaction in matter are known since a long time. However, their potential as a probe to give information on the interior of large structures is more recent, and only in the last years the literature has seen a large number of applications  made in this field, also due to the development of appropriate detectors, electronics, reconstruction and simulation algorithms. For a review of muon imaging techniques and applications, see \cite{Procureur2018}.

In this framework, the MEV (Muography of Etna Volcano) project brings together the experience of experimental physicists, engineers, geophysicists and volcanologists. In the present work we report the design, construction and testing of a high-resolution, large area muon telescope which consists of three position sensitive planes to track particles that cross the detector. 

\section{The MEV muon-tracking detector}
A muon tracking detector (telescope), able to measure the direction of the muons in its field of view (FOV), provides an experimental measurement of the particles flux, along different directions. Once the actual traversed thickness is known and taken into account for any specific direction, the measured flux can be used to calculate a density map of the target object. In principle, the combined use of several detectors, pointing to the object from different orientations, may produce a 3D density map of the object. A standard set-up for muon absorption experiments requires a telescope, usually employed in transmission mode (i.e. with no other object located behind the target, which is the unique responsible of the attenuation of the muon flux). The reconstruction of a large number of tracks in the FOV of the telescope allows the acquisition of a 2D radiographic map, with a resolution depending on the geometric characteristics of the telescope and its distance from the target. Many other aspects affect the detector performance, including its overall detection efficiency, response uniformity, sensitive area, alignment properties, duty cycle, cost and transportability.

\subsection{Mechanical construction and power source}
\begin{figure}[hbtp]
\centering
\includegraphics[width=0.9\columnwidth]{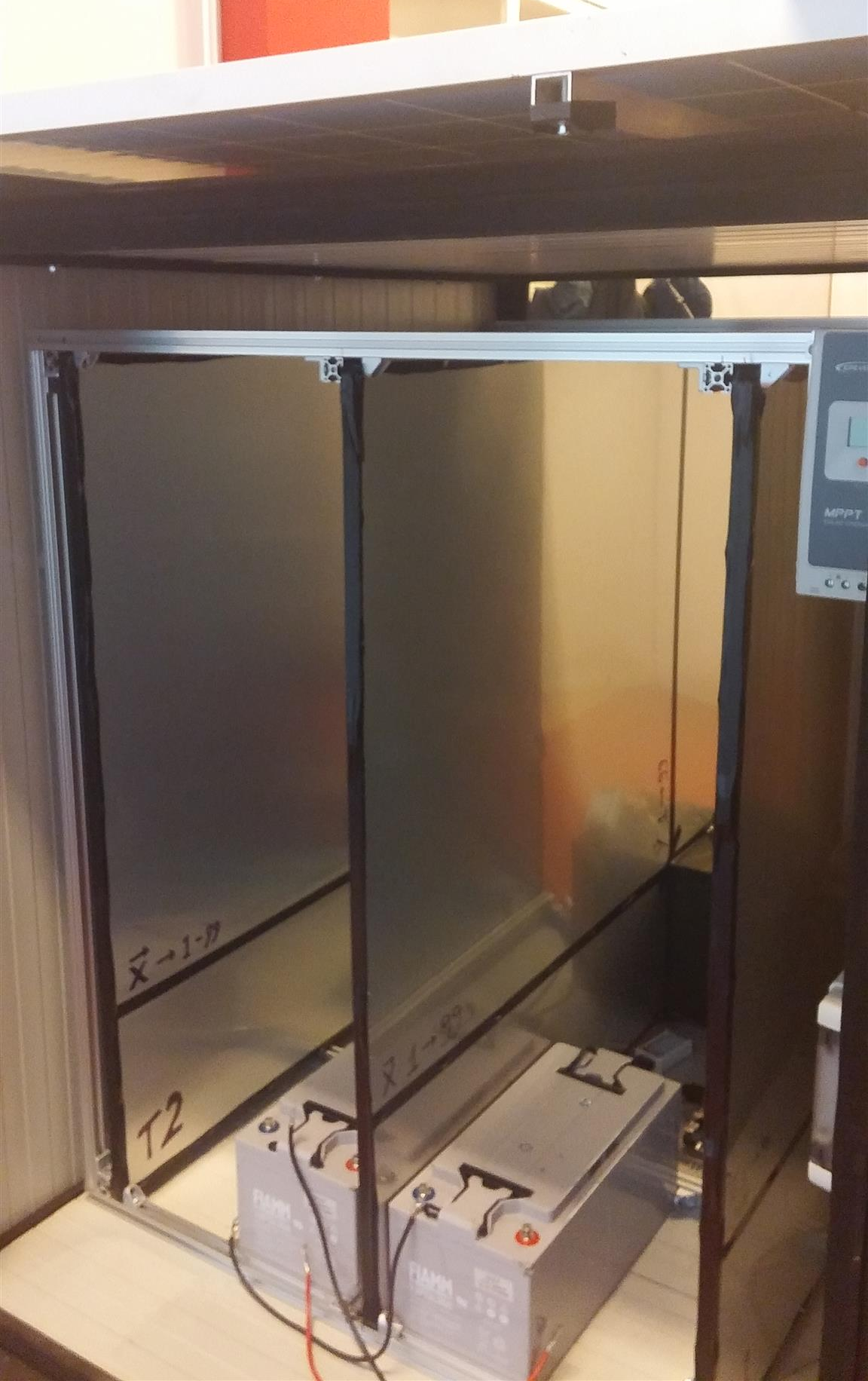}
\caption{\label{pic2}Assembly phase of the telescope. It is possible to see the three sensitive planes fixed vertically on an aluminium frame to keep the alignment inside the external box.}
\end{figure}

The first telescope developed for the MEV project was built at the Department of Physics of the University of Catania and was funded by FIR2014 (Future in Research) and strongly supported by the Parco dell’Etna national park and the Istituto Nazionale di Geofisica e Vulcanologia (INGV). The detector is based on three XY position-sensitive (PS) planes, with a sensitive area of \SI{1}{\square\meter}. The sensitive modules are enclosed in a cubic box with external side of about \SI{1.5}{\meter}, made with panels with double metallic cover and an inner isolating filling of polyurethane (see figure \ref{pic2}).

\begin{figure}[hbtp]
\centering
\includegraphics[width=\columnwidth]{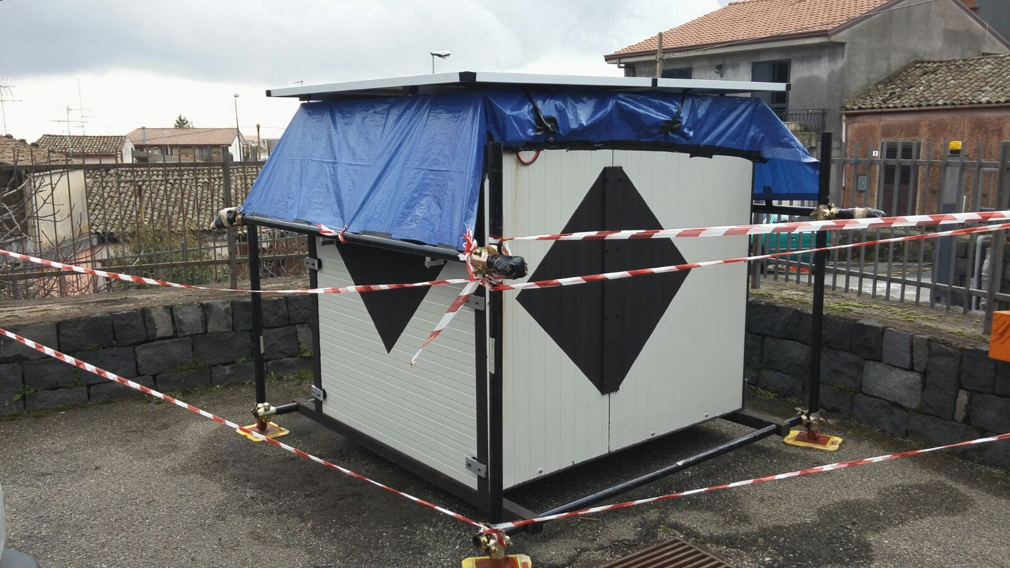}
\caption{\label{pic3}The telescope placed at the INGV facility in Nicolosi during the test phase.}
\end{figure}

Two solar panel (size: $\SI{150}{\centi\meter} \times \SI{80}{\centi\meter}$; peak power: \SI{260}{\watt}; output voltage: \SI{12}{\volt}) are mounted on the upper side of the box and charge two batteries (\SI{12}{\volt}, \SI{205}{\ampere\hour}) housed inside the box. The box is mounted on a modular frame made with scaffolding pipes which facilitate the transportation of the structure using a truck with a mechanical arm. This allows to transport the telescope already mounted to the measurement site. However, the internal aluminium structure which holds the PS planes is modular and can be assembled in the field. The frame lies on adjustable legs to cope with uneven terrain. Figure \ref{pic3} shows the telescope installed at the facility of INGV in the village of Nicolosi during the test phase.
The mechanical and power design of the detector makes it able to work out of the laboratory with no need of external power, in view of its future installation in the summit zone of Etna Volcano.

\subsection{The tracking modules}
Different detection system - emulsions \cite{Tanaka2007}, micromegas \cite{Morishima2017}, scintillators \cite{Tanaka2014,Marteau2017,Saracino2017}, multi-wire proportional chambers \cite{Olah2018} - have been employed in the construction of a muon telescope, but matrices made with scintillator strips are the best choice to do experiment on volcanoes. In fact, they do not high voltages, do not require any maintenance and the signal induced by the particles can be accessed in real-time. This choice was also guided by the consolidated experience acquired by several members of the project, e.g. \cite{Riggi2017,LoPresti2016,Carbone2014}.

The telescope can be thought as an imaginary parallelepiped where two PS planes are placed on opposite faces and the third detection matrix lies in the middle. Each PS plane consists of two layers of 99 Amcrys extruded plastic scintillator bars \cite{amcrys} (namely {\num{1 x 1 x 100} \si{\cubic\centi\meter}) with a central \SI{2.5}{\milli\meter} hole inside through which two \SI{1}{\milli\meter} Wave Length Shifting (WLS) fibres Saint-Gobain BCF-92 \cite{BCF92} are embedded to transport the photons to  a Position Sensitive PhotoMultiplier (PSPM). The plastic scintillator is coated with a white reflector on each external side. Figure \ref{pic0} shows some photos taken during the telescope assembling.

\begin{figure}[hbtp]
	\centering%
    \subfigure[\label{pic0a}]%
      {\includegraphics[width=\columnwidth]{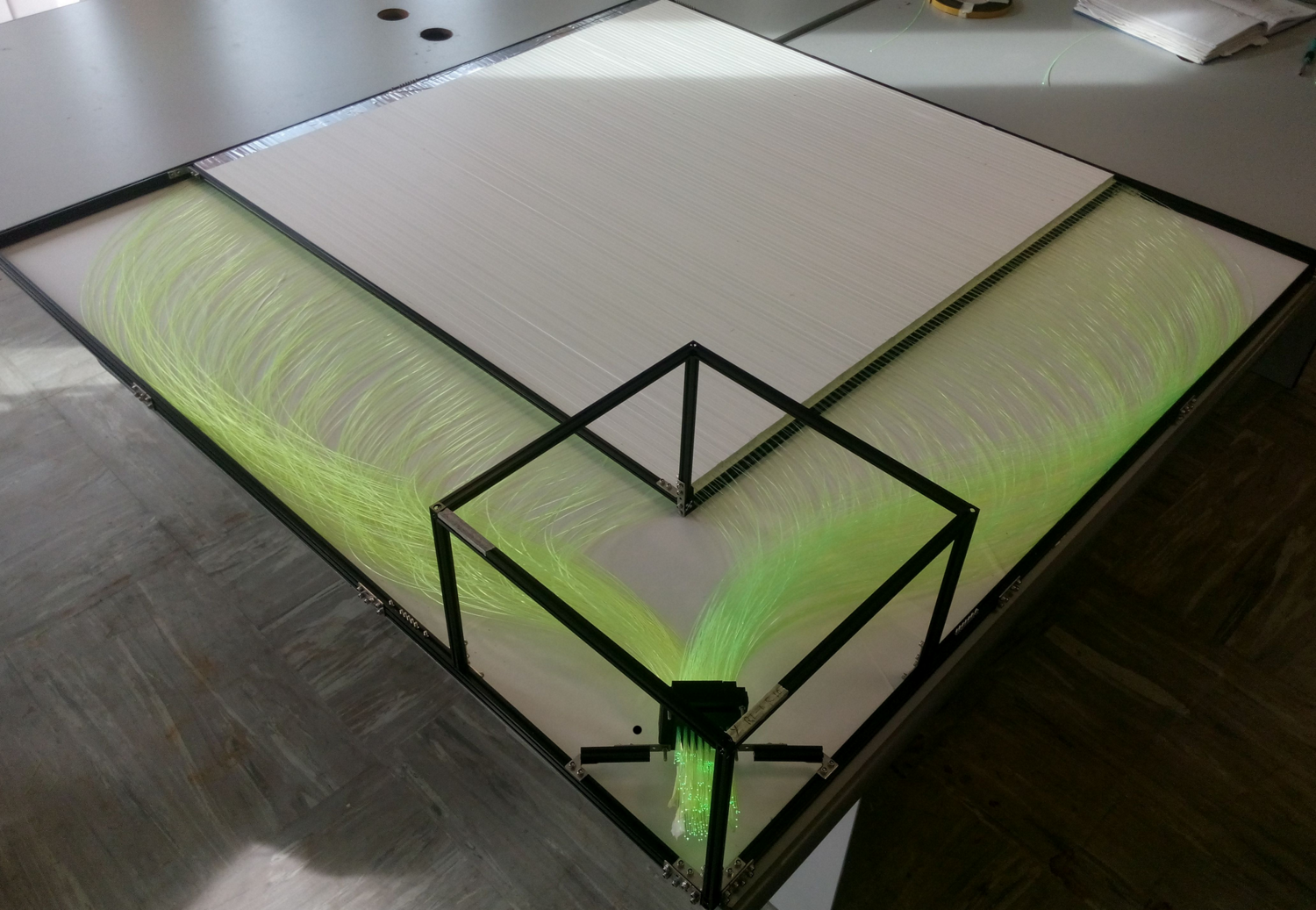}}\quad
	\subfigure[\label{pic0b}]%
      {\includegraphics[width=\columnwidth]{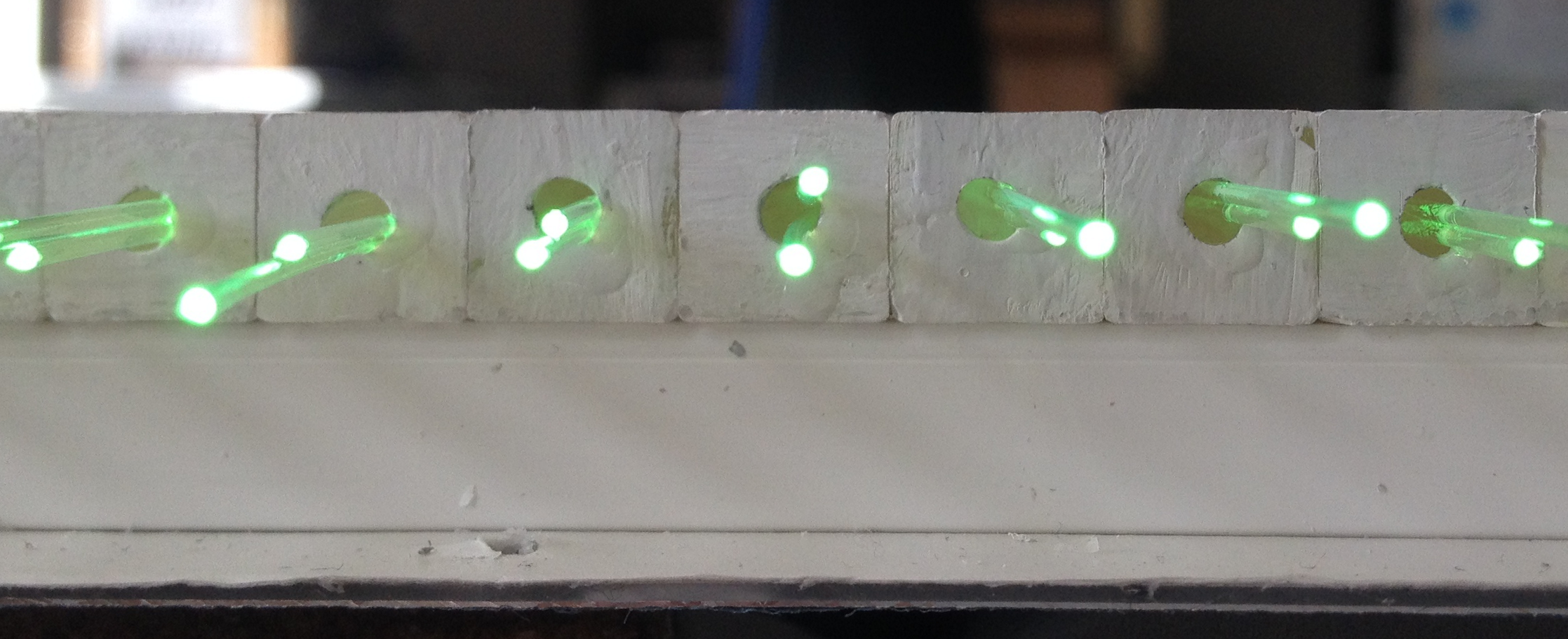}}
\caption{\label{pic0}Assembling of a telescope sensitive plane. In \ref{pic0a} it is possible to see the two layers of extruded plastic scintillator bars of a plane enclosed in a frame made with aluminium profiles; the \ref{pic0b} shows a detail of the two WLS fibres embedded inside each bar.}
\end{figure}

Each PS plane, therefore, is a matrix of $N \times N$ pixels, with $N = 99$. Let $f_{i,j}$ and $b_{k,l}$ be the pixels of front and back matrix, respectively, with $i,j,k$ and $l$ ranging from 1 to 99. The combination of all possible pixel pairs of outer matrices defines a set of $(2N-1)^2$ discrete directions of sight $\mathbf{r}_{m,n}$, with $m=i-k$, $n=j-l$. The direction $r_{0,0}$ is normal to the matrices and is parallel to the axis of the telescope, oriented from back (open-sky side) to front (target object side) and passing through its centre. The total solid angle covered by the telescope and its angular resolution depend on the number of pixels, their size $p$ and the distance $D$ separating front and back matrices. Figure \ref{fig1a} shows the angular resolution of our telescope with $p=100/N\,\si{\centi\meter}$ and $D=\SI{97}{\centi\meter}$. The angular aperture is about $\pm\SI{45}{\degree}$ and the resolution does not exceeds \SI{4e-4}{\steradian}, approximately. The detection area depends on the number of pixel pairs which share the same direction and on $p$. The geometric characteristics of the telescope are summarized by the acceptance function $\mathcal{T}$, (Fig. 4b) given in \si{\square\centi\meter\steradian} as defined in \cite{Lesparre2010}, and shown in figure \ref{fig1b}. The maximum acceptance, corresponding to the normal direction, reaches about $\SI{1}{\square\centi\meter\steradian}$.

\begin{figure}[hbtp]
	\centering%
    \subfigure[\label{fig1a}]%
      {\includegraphics[width=0.9\columnwidth]{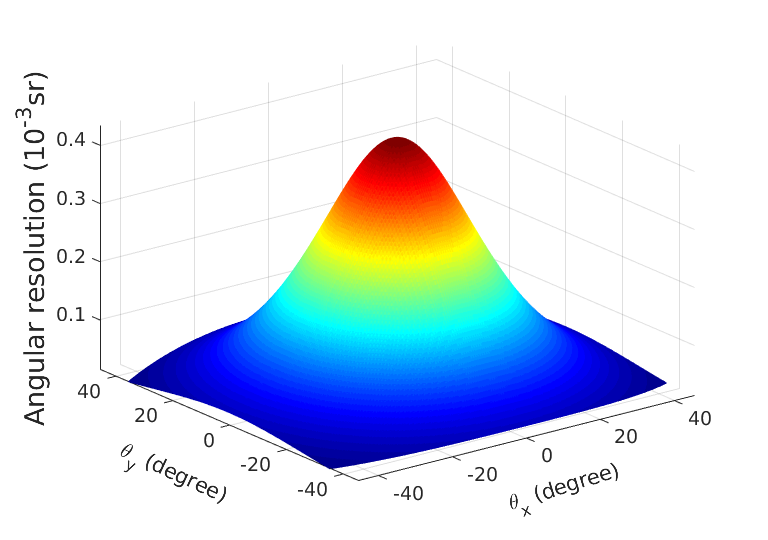}}\quad
	\subfigure[\label{fig1b}]%
      {\includegraphics[width=0.9\columnwidth]{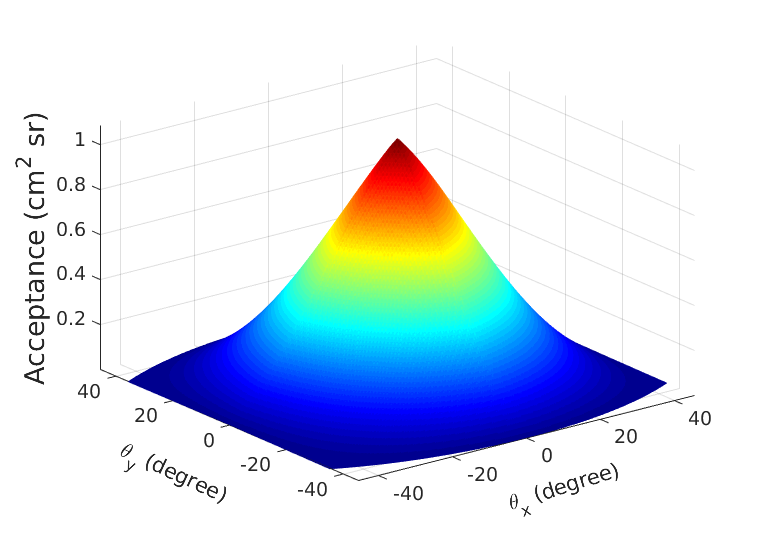}}
	\caption{\label{fig1}Distribution of the angular resolution (a) and the Acceptance $\mathcal{T}$ (b) for each discrete direction of sight of a telescope with three $99\times99$ matrices, $p\simeq\SI{1.01}{\centi\meter}$ and $D=\SI{97}{\centi\meter}$. Both quantities are displayed as functions of azimuth, $\theta_x$, and  zenith angle $\theta_y$. The telescope axis corresponds to $\theta_x = 0$, $\theta_y = 0$. The drawings were obtained using the software MATLAB.}
\end{figure}

\subsection{Front-End and Read-Out electronics}
The electronic chain is fully custom-designed for the purpose of the MEV project in order to have low-power consumption. It can be divided in two main level, the Front-End (FE) and the Read-Out (RO) electronic. The FE consists of three boards, one for each PS module. Each FE board houses a PSPM and a MAROC3 chip \cite{maroc}, which, in the pre-processing phase, pre-amplifies and shapes the analog signal from the PSPM with a peak time of about \SI{20}{\nano\second} and compares all the signals to a common threshold, giving a digital time-over threshold signal for each channel. The use of the PSPM guarantees a very low dark current  and greater stability of the response when temperature changes, compared to solid-state sensors (SiPM). A photo of a FE board is shown in figure \ref{pic1}.

\begin{figure}[hbtp]
\centering
\includegraphics[width=\columnwidth]{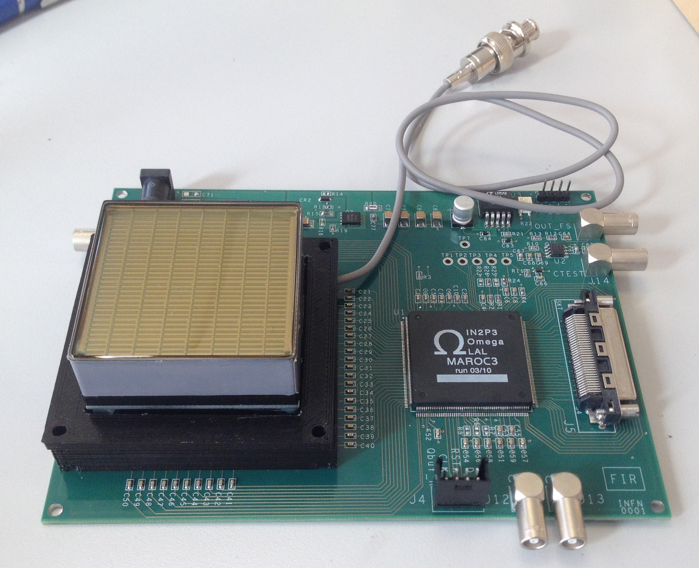}
\caption{\label{pic1}A FE board equipped with a Hamamatsu H8500 64 channels PSPM.}
\end{figure}

The output from each FE board is then acquired, filtered and processed by a National Instrument SOM (System-on-Module) \cite{som} on a single RO board, which houses also the sensing components for temperature and movement (accelerometer). A scheme of the complete electronic chain is reported in figure \ref{fig2} together with a photo of the RO board. The SOM allows a graphical approach for programming its FPGA with LabVIEW. A dedicated User Interface (UI) has been developed to manage and control all the parameters of the system. The data are sent by the LTE communication to a cloud storage at the Department of Physics. The power consumption of the whole telescope is about \SI{20}{\watt}, including the data transmission system.

\begin{figure}[hbtp]
\centering%
    \subfigure[\label{fig2a}]%
	 {\includegraphics[width=0.7\columnwidth]{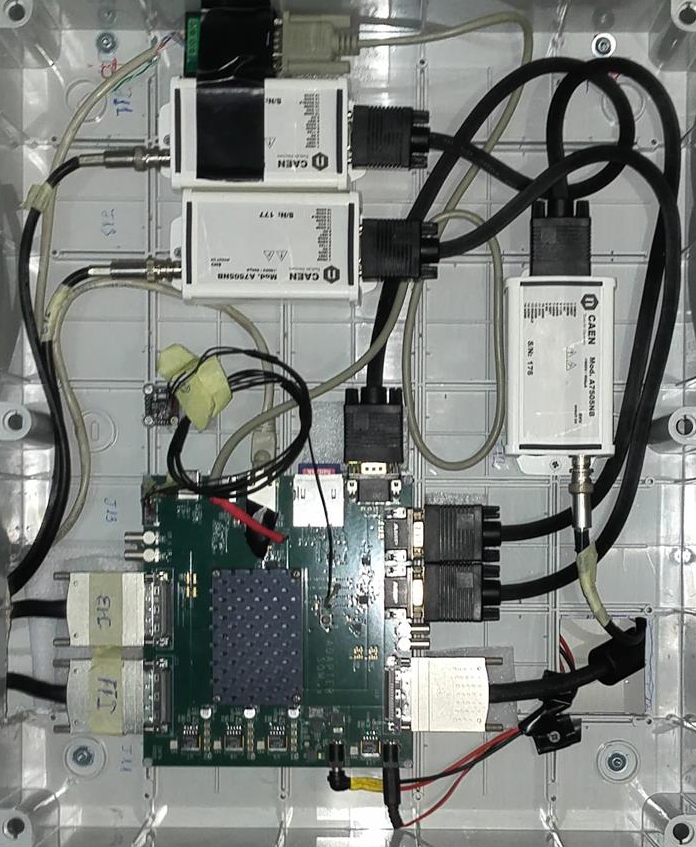}}
    \subfigure[\label{fig2b}]%
     {\includegraphics[width=0.9\columnwidth]{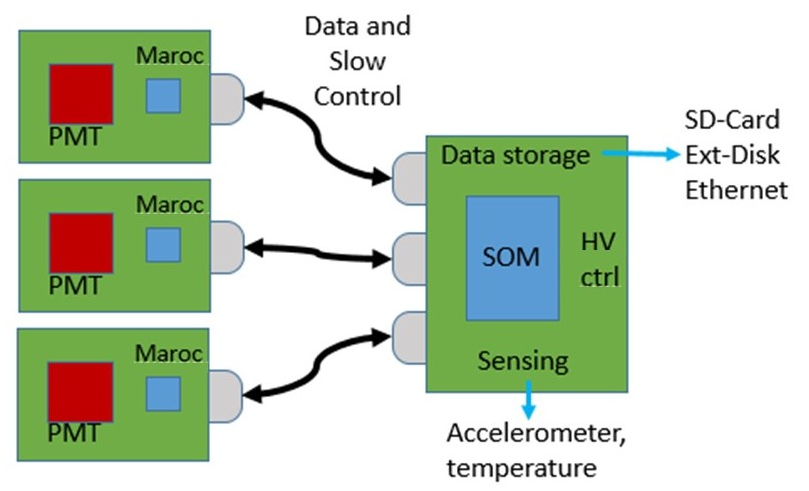}}
\caption{\label{fig2} (a) A photo of the RO board placed inside a plastic box with the modules for PSPMs power supply. (b) Diagram of the electronic chain of the telescope.}
\end{figure}

\subsection{Detection strategy and data acquisition}
The telescope exploits an innovative read-out architecture and trigger strategy, aimed at the optimization of detection efficiency and power consumption. Each PS plane has 198 intersecting strips, corresponding to $2\times198$ optical channels, assuming to separately read-out each of the two WLS per strip. A channel reduction \cite{LoPresti2015} system allows the minimization of the optical channels to be processed in order to reconstruct the impact coordinates, intended as the X and Y strips hit by the muon with maximum signal, in each plane. The routing of the WLS fibres according to the reduction system, ensure that just 40 channels per plane are sufficient. This makes it possible to couple the WLS fibres with a single Hamamatsu H8500 PSPM \cite{hamamatsuproductflyer}.

The MAROC3 chip compares each pre-processed signal to a remotely settable threshold, producing at its output an encoded digital representation of the detector response to the passage of a muon. A calibration procedure, repeated periodically, equalizes the channel efficiency, that is a combination of the gain and optical coupling of the single channel, in order to uniform the plane detection efficiency.

Another advantage of the channel reduction system is that it makes possible to use a single SOM to ensure the synchronization of the data coming from each FE board. The SOM samples simultaneously at 200 MHz  all the 120 channels, corresponding to the response of the three planes of the detector. The trigger condition for the acceptance of an event, a muon track in the FOV of the telescope, is the presence of a X and Y signal in each plane, within a time window of 80 ns. This trigger condition - three two-fold coincidences - filters out all the signal generated by the very low dark current of the PSPM. The samples related to an event, an array of $120\times16$ bit, is transferred to the cloud storage by LTE communication system.

In order to reduce to a minimum the spurious coincidences, an off-line analysis further filters the data and considers only the events which produce a fourth-fold coincidence in each plane inside the time window; then, the analysis checks that the recorded XY on each plane match a linear trajectory by means of a 3D linear fit and discards not straight tracks. 

\section{First results}
During the test phase, the telescope was installed at the INGV facility located in Nicolosi, a village \SI{13}{\kilo\meter} far from Catania. This phase started in January 2017 and ended in July of the same year, lasting about 120 days. The effective time of data acquisition was actually shorter (about 70 days), due to tuning and adjustments of the whole data acquisition system, in particular regarding the communication with network and data storage. 

\begin{figure}[hbtp]
\centering
\includegraphics[width=\columnwidth]{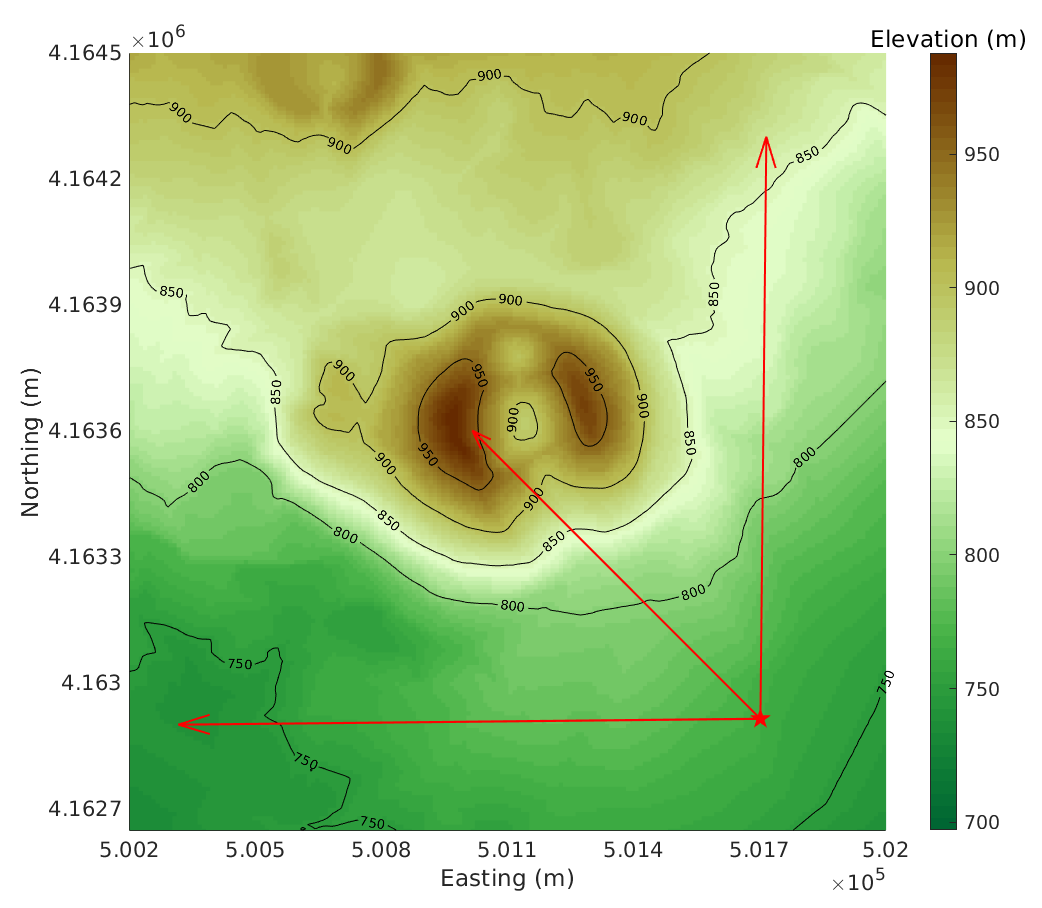}
\caption{\label{fig3a}Contour map of the Monti Rossi zone, obtained from a 10m Digital Elevation Model by INGV (2014). The red star marks the position of the telescope during the test phase and the arrows show its azimuthal field of view with the telescope axis in the middle.}
\end{figure}

The test phase involved the muography of Monti Rossi, an extinct volcanic cone, at a distance of about \SI{900}{\meter} from the telescope. The wrong plural denomination, “Red Mountains”, arises from the presence of two peaks that seems to be two distinct mountains, but are rather the rest of a single volcanic cone (fig. \ref{fig3a}). We did not expect to observe strong density anomalies inside the Monti Rossi, but the target shape, size and distance from the telescope are very similar to what expected for the final objective, the North-East crater of Etna Volcano. The result of this measurement is shown in figure \ref{fig3}. In the upper part of figure \ref{fig3} ($\Delta Y > 0$) the profile of the Monti Rossi is clearly visible. The rising of the muon counts as concentric rings is due to the convolution of telescope acceptance and muon flux dependence by zenith angle.

\begin{figure}[hbtp]
\centering
\includegraphics[width=\columnwidth]{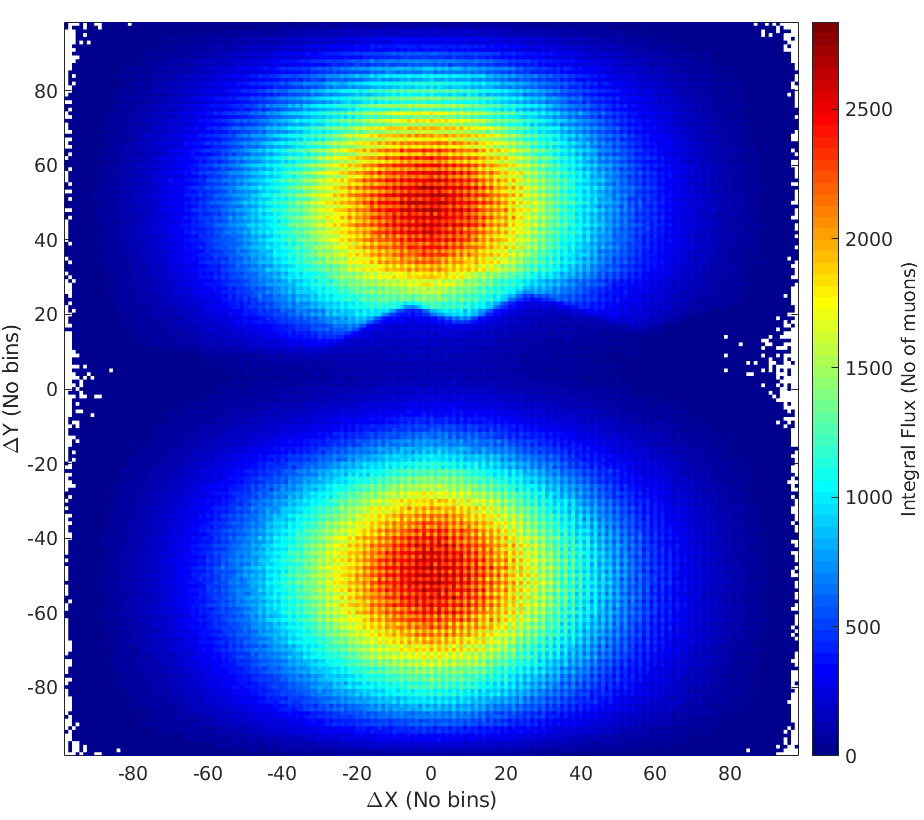}
\caption{\label{fig3}Integral of the muon flux acquired during the test phase (January - July 2017). The differences in X and Y coordinates of the crossing point of the track in the external planes are expressed in terms of bins or number of bars.}
\end{figure}

We follow the common practice to express muography data as a function of X and Y displacement, $\Delta X$ and $\Delta Y$, between the entrance and exit coordinates of the muon tracks in the telescope external planes. $\Delta Y > 0 $ identify muons coming from the front of the detector, i.e. from the side pointed towards the object to be imaged; $\Delta Y < 0 $ corresponds to muons entering the detector from the back side. In this way, all the tracks with the same $(\Delta X, \Delta Y)$, i.e parallel to the direction $r_{m,n}$, will be grouped together. This is not a simple convention to display muography data, but reflects the assumption that the telescope is considered as a point compared to the target object. If the telescope is oriented such that no object enters in the field of view for $\Delta Y < 0$, it is possible to acquire simultaneously the absolute muon flux which comes from open sky and the flux after traversing the object of interest.

In figure \ref{fig3} it is possible to notice a regular pattern that is superimposed over the image, a sort of cross-hatch, where one pixel every two, in both X and Y directions,  has a higher counts value. This is  a drawback in geometric efficiency, due to  presence of a third plane in the middle. As shown in figure \ref{fig4}, given a  difference in position between Y bars in external planes, e.g. two bars, a particle may either (i) cross the plane in the middle intersecting a bar nearly in its centre with a good path inside it (green line), or (ii) intersects the plane in the middle nearly in the edge between two adjacent bars (dotted red line). If (ii) is the case, the generated scintillation light could be too weak to be detected. This effect on the muon counts integral can be overtaken as explained in the following.

\begin{figure}[hbtp]
\centering
\includegraphics[width=\columnwidth]{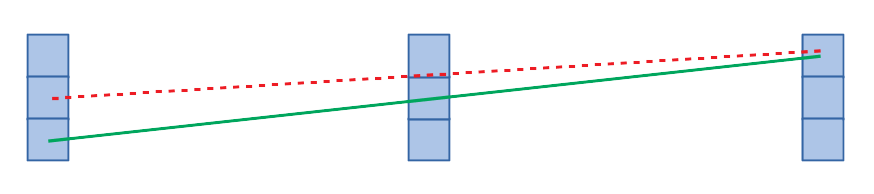}
\caption{\label{fig4}A sketch of the transversal section of the telescope with 3 bars for each plane in one direction. The green and dotted red lines give an example of two different events that explain the cross-hatch pattern in figure \ref{fig3}.}
\end{figure}

The telescope was thought to be placed horizontally, i.e. with its axis (corresponding to its principal line of sight $r_{0,0}$) oriented along zenith angle equal to \SI{90}{\degree}. In this way, the muon fluxes acquired from both sides of the detector could be compared easily because they span the same zenith angular range and a measure of muon transmission is directly accessible. Muon transmission is defined as the ratio between the muon flux reaching the detector after traversing the object under investigation divided by the flux impinging on the target surface before crossing it, that corresponds to the flux from open sky. We refer to reference \cite{Saracino2017} for an exhaustive definition of \textit{muon transmission}.

The measured muon transmission calculated as the ratio between measured flux for $\Delta Y > 0$ and $\Delta Y < 0$ is shown in figure \ref{fig6a}. The inner part of the cone is characterized by a muon transmission value which slightly decreases towards higher zenith angles and which do not reveal any anomaly clearly related to features of the inner structure of the crater. 

Figure \ref{fig6b} shows the traversed thickness for each direction in the same portion of the telescope FOV displayed in fig. \ref{fig6a}. Comparing the two figures, it is possible to notice that muon transmission decreases where thickness increases, as we could expect. For zenith angles near to the horizon the transmission augments, but we should expect no counts because the thickness becomes very large. This background noise affects all data obtained through this kind of muography application, regardless the technique for muon tracking.

\begin{figure}[hbtp]
	\centering
	\subfigure[\label{fig6a}]%
     {\includegraphics[width=\columnwidth]{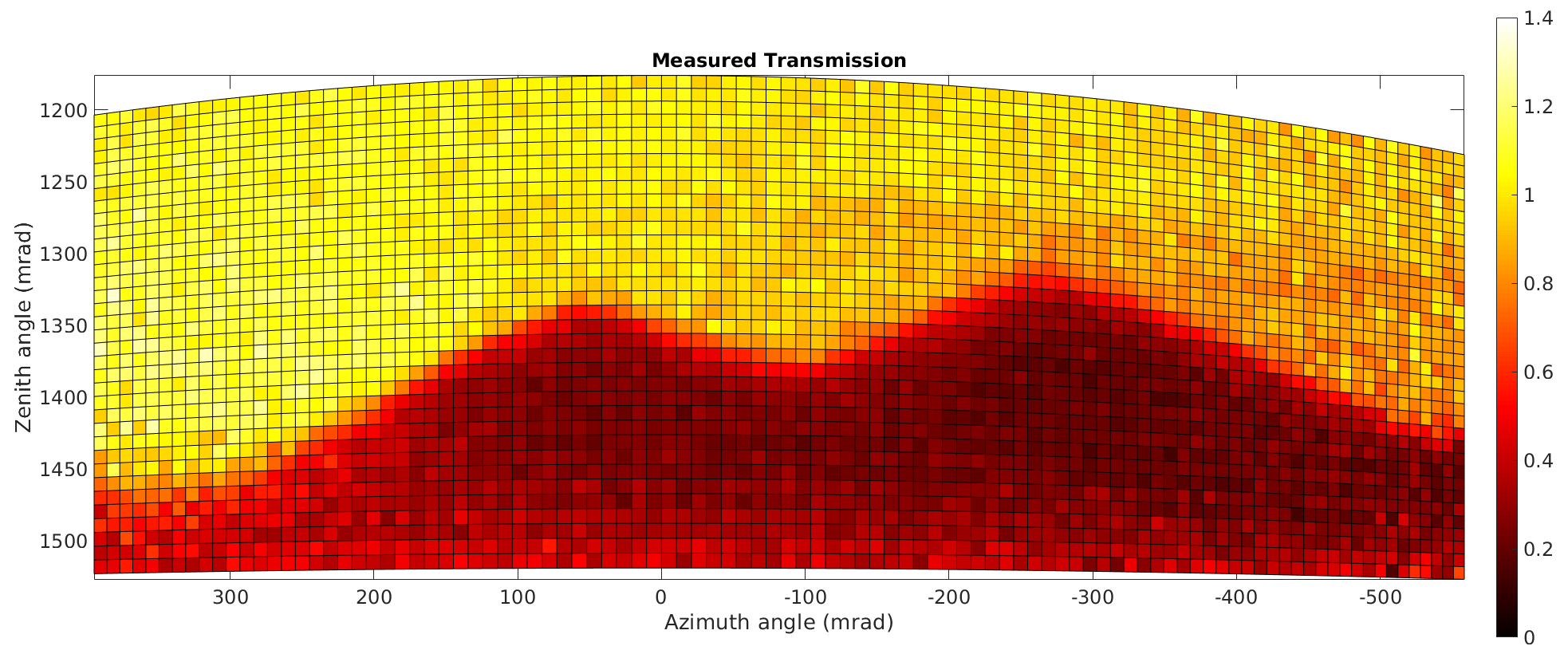}}\quad
    \subfigure[\label{fig6b}]%
     {\includegraphics[width=\columnwidth]{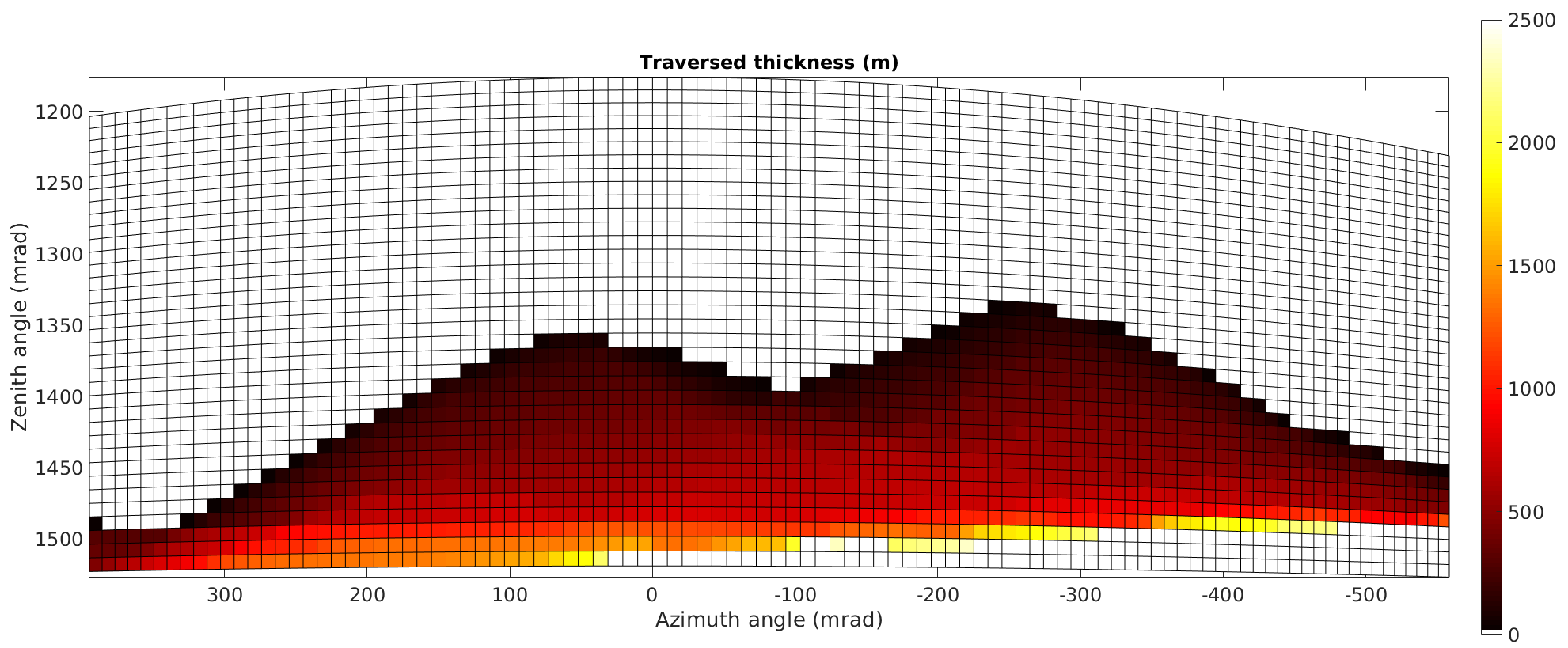}}
    \subfigure[\label{fig6c}]%
     {\includegraphics[width=0.8\columnwidth]{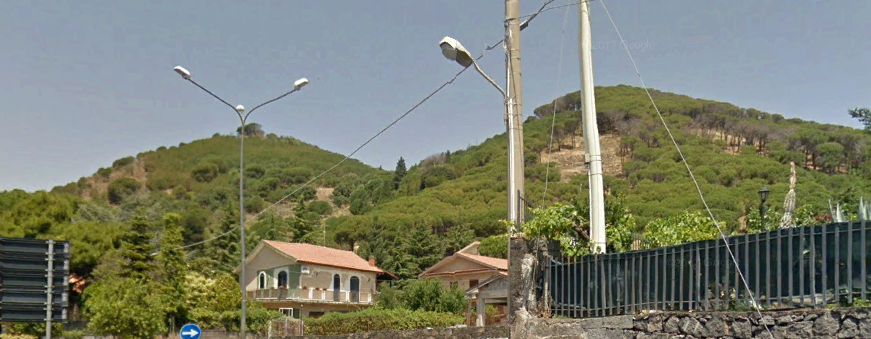}}
\caption{\label{fig5}Comparison between measured muon transmission for each trajectory given by zenith and azimuth angles, traversed thickness calculated from DEM and a picture of Monti Rossi from a point close to the installation site of the telescope. The trajectory with zenith angle = \SI{\pi/2}{\radian} and azimuth angle = 0 corresponds to the telescope axis.}
\end{figure}

Although the origin of the background particles is not fully understood, some prior works considered possible candidates. For instance, Jourde et al. \cite{Jourde2013} proposed that it arises from particles entering the back of the detector with upward-going trajectories after being scattered by the soil. These particles mimic the signals of penetrating muons, since upward-going particles have trajectories identical to those of downward-going muons emerging from target mountains. They have detected these upward-going particles in several locations by dedicated Time-of-Flight (TOF) analysis on data from scintillation detectors. Although the upward-going particles are part of the source of the background particles, they do not explain all the excess in the particle flux. 

Nishiyama et al. \cite{Nishiyama2014} proposed that the background particles are due to low-energy charged particles ($E\leq \SI{1}{\GeV}$) which are scattered onto detectors from random directions. They installed two emulsion detectors with different energy thresholds (0.1 and 1.0 \si{\GeV}) at the foot of a small lava dome (Mt. Showa-Shinzan) and demonstrated that only the detector with the higher energy threshold yields appropriate particle flux values for the lava. However, the origins of these low-energy particles have not been verified, neither their connection with the upward-going particles.

Another difference between figures \ref{fig6a} and \ref{fig6b} is that in the measured transmission map, the profile of the mountain seems to be blurred. This is due to the fact that the target object itself behaves as a source of background that produces a sort of halo, as a consequence of the multiple scattering that muons experienced going through it.

The last panel of figure \ref{fig5} shows a picture of the Monti Rossi taken from a point near the telescope location at Nicolosi. The target object was not directly visible from installation site, because of surrounding buildings. Nevertheless, it is clearly visible that the muographic image reproduce very well the shape of the crater shown in the photograph and in the thickness map (Fig. 10b), obtained from the Digital Elevation Model (DEM).

\section{Conclusions and outlooks}
The results presented in this paper confirm the success of the test phase of the new muography detector developed under the MEV project. The mechanical structure that contains the three sensitive planes proved to be relatively light and water tight. All the problems related to wireless communication with the detector were solved and the telescope is fully remotely controllable and independent from external power sources. The first muographic image of a geophysical object was acquired with a promising quality. Hence, we decided to transfer the telescope to the summit zone of Etna Volcano, with the aim of imaging the active North-East crater.

In order to address background noise issue, we plan to add a module to measure the TOF of the particle between the external planes. This will reduce the influence of bad reconstructed tracks near the horizon. The TOF module is under test in the laboratory and will be integrated into the telescope as soon as possible.

The scattering due to to the target object itself is practically unavoidable. Other projects, e.g. \cite{Olah2018}, have employed planes of high Z material, mainly lead, in order to suppress the tracks due to low energy particles and consider only muons that go straight through the structure to be imaged. However, this approach is not feasible in the summit zone of a tall volcano like Etna, where limited access due to harsh conditions prevents the use of heavy equipment. Furthermore, a suitably designed detector is required, able to recognize the internal scattering due to high Z material.

\section*{}
\bibliographystyle{elsarticle-num} 
\bibliography{BibTeX/AMCRYS.bib,BibTeX/Allard.bib,BibTeX/BCF92.bib,BibTeX/CFerlito_2018.bib,BibTeX/ChouetMatoza2013.bib,BibTeX/DCarbone2013.bib,BibTeX/DCarbone2017.bib,BibTeX/Dzurisin2007.bib,BibTeX/Gilbert2008.bib,BibTeX/Gonnermann.bib,BibTeX/H8500.bib,BibTeX/Jourde.bib,BibTeX/Lesparre2010.bib,BibTeX/LoPresti_2015.bib,BibTeX/LoPresti_2016.bib,BibTeX/MAROC.bib,BibTeX/Marteau2017.bib,BibTeX/Morishima_2017.bib,BibTeX/Nishiyama2014.bib,BibTeX/Olah_2018.bib,BibTeX/Procureur2018.bib,BibTeX/Riggi2017.bib,BibTeX/Saracino_2017.bib,BibTeX/SOM.bib,BibTeX/Tanaka2007.bib,BibTeX/Tanaka2014.bib}

\end{document}